\documentstyle[12pt,epsf]{article}
\advance\textheight\topskip
\begin{document}
\baselineskip=6mm
\newcommand{\rf}[1]{(\ref{#1})}

\def\GS{G_{\mbox{\tiny S}}}
\def\GV{G_{\mbox{\tiny V}}}
\def\LC{\lambda_{\mbox{\tiny c}}}
\def\MF{m_{\mbox{\tiny f}}}

\begin{titlepage}
\begin{flushright}
KANAZAWA/96-23\\
GEF-TH/96-18
\end{flushright}
\quad\\
\vspace{1.8cm}
\begin{center}
{\bf\LARGE Non-Perturbative Renormalization Group Analysis}\\
\medskip
{\bf\LARGE of the Chiral Critical Behaviors in QED}\\
\vspace{1.3cm}
Ken-Ichi Aoki
\footnote{e-mail address: aoki@hep.s.kanazawa-u.ac.jp}, 
Keiichi Morikawa, Jun-Ichi Sumi, 
Haruhiko Terao$^{*}$
\footnote{On leave of absence from the Department of Physics,
Kanazawa University,\\\null\hskip7mm e-mail address: terao@dipfis.ge.infn.it}
\\
\vspace{0.2cm}
and Masashi Tomoyose\\
\bigskip
\vspace{0.8cm}
Department of Physics, Kanazawa University,\\
\vspace{0.0cm}
Kakuma-machi, Kanazawa 920-11, Japan\\
\vspace{1cm}
$^*$Dipartiment di Fisica - Universit\`{a} di Genova,\\
\vspace{0.0cm}
Istituto Nazionale di Fisica Nucleare - sez. di Genova,\\
\vspace{0.0cm}
Via Dodecaneso, 33 - 16146 Genova, Italy

\vspace{1cm}

{\bf Abstract}
\end{center}
We study the chiral critical behaviors of QED by using 
Non-Perturbative Renormalization Group (NPRG). 
Taking account of the non-ladder contributions, our flow equations 
are free from the gauge parameter ($\alpha$) dependence.
We clarify the chiral phase structure, 
and calculate the anomalous dimension of $\bar\psi\psi$,  which is
enhanced compared to the ladder approximation.  
We find that the cutoff scheme dependence of the physical results is
very small.
\end{titlepage}
%
%
\section{Introduction}

\hskip\parindent The dynamical chiral symmetry breaking in gauge theories is 
one of the most important subjects in particle physics. 
Many efforts have been devoted to this problem particularly by solving
the Schwinger-Dyson (SD) equation for the fermion self-energy mainly
in the ladder approximation with the Landau gauge\cite{SDmain}. 
Adding the four-fermi interactions, the chiral phase diagram and the
critical behaviors have been investigated\cite{SDfourfermi}
and its characteristic
structures with the large anomalous dimension of $\bar\psi\psi$
operator has been applied to various models beyond the standard 
model\cite{SDmodel,topcondensation}.
In QCD, the ladder SD with the running gauge coupling constant, 
the improved ladder, gives good results even quantitatively\cite{SDimproved}. 
However
the ladder SD is unsatisfactory in many aspects including the strong 
gauge dependence\cite{SDdifficulty}, 
the difficulty to proceed beyond the ladder\cite{SDbeyond}, no firm base
for the improved ladder. 

Quite recently, numerical analysis with the
Non-Perturbative Renormalization Group (NPRG) \cite{WILSON,WETTERICH} 
draws much attention as a promising new method to study 
non-perturbative phenomena of the field theory. 
One of the essential features of the method is that it does not
employ any series expansion and therefore is expected to give
converging results when improving the approximation, which shows a 
great contrast against other non-perturbative expansion 
methods such as $1/N$ or $\epsilon$ expansions \cite{souma,comoving}.

Here we briefly explain the general aspects of the NPRG approach. 
The RG equation is the nonlinear functional differential equation 
for the Wilsonian effective action, the exact form of which is explicitly 
written down. To approximate the RG equation, 
we project it onto a small dimensional subspace of the original 
full theory space. 
To improve the approximation, we enlarge the subspace step by step.
As the lowest order, we take the local potential approximation (LPA)\cite{lpa}.
The sharp cutoff LPA NPRG has been applied to 
the dynamical chiral symmetry breaking in gauge theories and it
successfully solves the critical behaviors \cite{terao,teraow}.
When we further artificially truncate the LPA down to its 
`ladder parts', it reproduces the Landau gauge ladder SD results, 
even the improved ladder results when
we adopt the running gauge coupling constant.
Thus the LPA NPRG should be regarded as a non-ladder improvement of
the ladder SD results, in the course of the systematic approximation
of NPRG.

In this article we proceed beyond the LPA analysis of the dynamical
chiral symmetry breaking in gauge theories including anomalous
dimensions of fields, which brings about two new issues to be carefully
dealt with.
First, we have to adopt a smooth cutoff scheme
to avoid the sharp cutoff singularities for derivative interactions.
What we call the cutoff scheme is the profile of a function to suppress the propagation 
of the low energy modes in the internal lines and is introduced through the 
momentum dependent mass terms in the Lagrangian, 
\begin{equation}
\frac{1}{2}\int \frac{d^4 q}{(2\pi)^4}\phi(-{\bf q})\cdot
\Lambda^2\cdot C(\Lambda^2/q^2)\cdot\phi({\bf q}),
\end{equation} 
where function $C$ defines the cutoff scheme.
In the full space NPRG, physical results do not depend on the cutoff
scheme. However the subspace approximation breaks this
independence\cite{SCHEME}, and the cutoff scheme dependence
must be checked explicitly.

Second, 
the NPRG is not manifestly compatible with the gauge symmetries due to 
the momentum cutoff. Therefore we have to introduce 
the gauge non-invariant operators to our subspace and constrain them
by certain identity, so-called `Modified Ward-Takahashi-Slavnov-Taylor 
Identity' \cite{MSTI}. 
This identity ensures the gauge invariance of the total solutions of NPRG.

In the following sections, we formulate the NPRG flow equations
beyond the LPA and calculate critical behaviors in QED.
Further we investigate the cutoff scheme dependence and 
the gauge dependence in the covariant $\alpha$-gauge.
These first results beyond the  LPA may give us a
clue to the total reliability of the NPRG results.

As mentioned above, the ladder SD results have been reproduced in 
the LPA by further truncating the flow equations.
Note that this truncation cannot be regarded as any projection of the 
RG equation to a subspace. This is crucial for the gauge independence.
Therefore the ladder truncation necessarily breaks the gauge invariance.
In our system, e.g., the crossed ladder diagrams 
are included as well as the usual ladder diagrams, keeping the
gauge independence of the flows (in some special cutoff scheme), while
the ladder SD takes only the ladder parts, and thus suffers from disastrous
gauge dependence \cite{SDdifficulty}. 
%
%
%
\section{RG flow equation for the massless QED}

\hskip\parindent 
We start with the evolution equation for the effective action 
$\Gamma_\Lambda[A_\mu,\bar{\psi},\psi]$\cite{WETTERICH}, 
where $\Lambda$ is the infra-red momentum cutoff for the quantum corrections.
The effective action is nothing but the one 
particle irreducible parts of the Wilsonian effective action. 
The infrared cutoff is introduced by 
adding the following terms to the bare action: 
\begin{equation}
\Delta S_{cut}=
\int \frac{d^4 q}{(2\pi)^4}\left \{
\frac{1}{2} A_\mu(-{\bf q})\cdot
\Lambda^2\cdot(\Lambda/q)^{2k}\cdot A_\mu({\bf q})+
(\Lambda/q)^{2k+2}\cdot q_\mu\cdot
\bar{\psi}(-{\bf q})\gamma_\mu\psi({\bf q})
\right \},
\end{equation}
where a parameter $k$ labels the cutoff scheme and 
we take the values $k=1,2,3,\cdots$.
With larger $k$ the cutoff profile becomes sharper and it reaches the
$\theta$-function cutoff at $k=\infty$. 
Since the above curoff terms maintain the chiral symmetry, 
the effective action $\Gamma_\Lambda$ also respects it. 
The evolution equation with this cutoff scheme is written as, 
\begin{equation}
\Lambda\frac{\partial}{\partial\Lambda}\Gamma_\Lambda=
(k+1)\cdot{\rm Str}\left \{
{\bf C}^{-1}\cdot\left(
{\bf C}^{-1}+
\frac{\overrightarrow{\delta}}{\delta\Phi^T}
\Gamma_\Lambda
\frac{\overleftarrow{\delta}}{\delta\Phi}
\right)^{-1}\right\} ,
\label{RGwett}
\end{equation}
where $\Phi$ is a shorthand notation of 
the fields $\Phi=\{A_\mu,\psi,\bar{\psi}\}$ and ${\bf C}^{-1}$ is the matrix,
\begin{equation}
{\bf C}^{-1}({\bf q})\equiv
\left(\begin{array}{c|c}
\Lambda^2\cdot(\Lambda/q)^{2k}\cdot\delta_{\mu\nu} & 
{\bf0} \\
      &      \\
\noalign{\vskip-3mm}
\hline
      &      \\
\noalign{\vskip-3mm}
{\bf 0} & \begin{array}{cc}
   0 & (\Lambda/q)^{2k+2}\cdot q_\mu\cdot\gamma_\mu^T \\
   (\Lambda/q)^{2k+2}\cdot q_\mu\cdot\gamma_\mu  & 0
\end{array}
\end{array}
\right)\; .
\end{equation}

To approximate the RG equation (\ref{RGwett}), 
we project it onto a finite dimensional subspace. 
Specifying a subspace defines an approximation.
Here we take a subspace spanned by the seven chiral invariant operators, 
\begin{eqnarray}
\Gamma_\Lambda[A_\mu,\bar{\psi},\psi]&=&\int d^4x\left\{
\frac{1}{4}Z_3 F_{\mu\nu}^2+\frac{1}{2\alpha}(\partial\cdot A)^2
+\frac{1}{2}m^2\Lambda^2 A_\mu^2+
Z_1\bar{\psi}\gamma_\mu i\partial_\mu\psi
+e A_\mu\bar{\psi}\gamma_\mu\psi\right.\nonumber\\
&&\left.-\frac{1}{2}\frac{\GS}{\Lambda^2}
\left[
(\bar{\psi}\psi)^2-(\bar{\psi}\gamma_5\psi)^2
\right]
-\frac{1}{2}\frac{\GV}{\Lambda^2}
\left[
(\bar{\psi}\gamma_\mu\psi)^2+(\bar{\psi}\gamma_5\gamma_\mu\psi)^2
\right]
\right\}\label{effective}\; ,
\end{eqnarray}
where $F_{\mu\nu}$ is the field strength and $m^2$ is the photon mass 
which compensates for the lack of the gauge invariance of 
$\Gamma_\Lambda$.

Note that the chiral and parity invariant 4-fermi interactions
without derivatives are 
spanned by the above two independent operators $\GS, \GV$.
Within the non-derivative (local potential) approximation 
for multi-fermi operators, the 4-fermi flow equations are not 
affected at all by other higher dimensional multi-fermi operators,
which are composed of at least 8-fermi operators due to the chiral
symmetry. In such case, the 4-fermi flows completely determine the
critical behaviors for the local potential approximated multi-fermi system,
although higher multi-fermi operator flows are necessary when
calculating the dynamically generated fermion mass and the chiral
condensates \cite{teraow}.

Substituting the effective action (\ref{effective}) to Eq.(\ref{RGwett}), 
we obtain the RG flow equations for the 
dimensionless coupling constants $i.e.$ $Z_3,Z_1,\alpha,m^2,e,G_S,G_V$
(see Fig.\ref{feynman1}). 

\vskip5mm
\refstepcounter{figure}
\label{feynman1}
\centerline{\epsfxsize=16cm\epsfbox{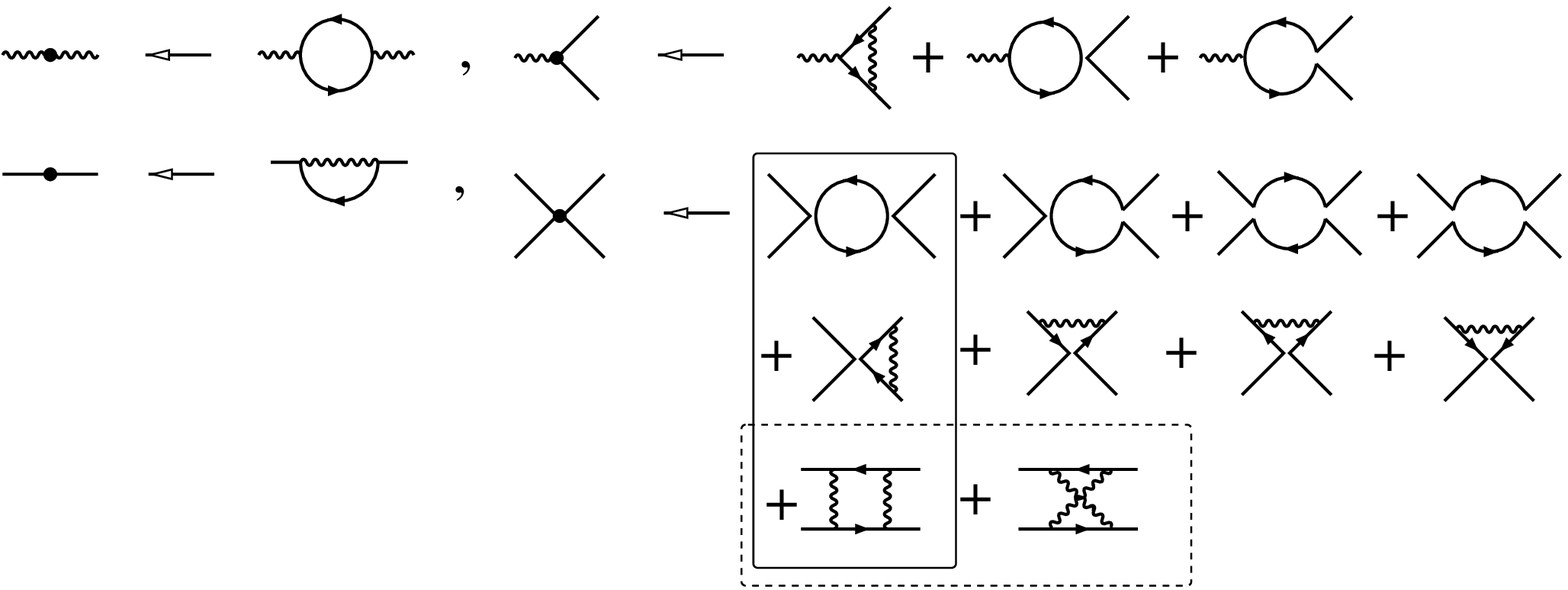}}
\vskip5mm
\centerline{\parbox{0.9\textwidth}{{\footnotesize
Figure \ref{feynman1}:
Diagrams incorporated in our approximation. The diagrams in the 
solid box correspond to the ladder part and the diagrams in the dashed box 
are crucial for the gauge invariance. 
}}}
\vskip8mm

The effective action (\ref{effective}) actually contains 
two redundant parameters due to the wave function renormalization. 
We set $Z_1$ and $ Z_3$ to be unity by 
rescaling fields $A_\mu,\psi,\bar{\psi}$. 
Consequently, the RG equations read,
\begin{eqnarray}
&\displaystyle\frac{\partial}{\partial t}m^2&=
\left(2-\frac{e^2}{6\pi^2}\right)m^2
-4I_1(k)e^2,
\label{mass}\\
&\displaystyle\frac{\partial}{\partial t}e&=
-\frac{1}{2}\left(3I_4(k,m^2)-I_5(k,m^2,\alpha)\right)e^3
-\frac{1}{12\pi^2}e^3
+2e\left(G_S+4G_V\right)I_1(k),
\label{gauge}\\
&\displaystyle\frac{\partial}{\partial t}G_S&=
-2G_S
-\left(3I_4(k,m^2)-I_5(k,m^2,\alpha)\right)e^2G_S
+6I_3(k,m^2)e^2G_S
\nonumber\\
&&\qquad\qquad
+2I_1(k)G_S(3G_S+8G_V)
-6I_2(k,m^2)e^4,
\label{scalar}\\
&\displaystyle\frac{\partial}{\partial t}G_V&=
-2G_V
-6I_3(k,m^2)e^2G_V
-\left(3I_4(k,m^2)-I_5(k,m^2,\alpha)\right)e^2G_V
\nonumber\\
&&\qquad\qquad
+I_1(k)G_S^2
-3I_2(k,m^2)e^4,
\label{vector}\\
&\displaystyle\frac{\partial}{\partial t}\alpha&=
\frac{1}{6\pi^2}e^2\alpha.
\label{alpha}
\end{eqnarray}
where $t$ is a cutoff scale parameter defined by
$\Lambda=\Lambda_0{\rm e}^{-t}$. 
There appear five independent functions of $k,\alpha$ and $m^2$.
These functions are defined by the following integrals:
\begin{eqnarray}
I_1(k)&=&
\frac{k+1}{8\pi^2}
\int_0^\infty dxx^{2k+2}
\left(\frac{1}{1+x^{k+1}}\right)^3\; ,\\
I_2(k,m^2)&=&
\frac{k+1}{8\pi^2}
\int_0^\infty dxx^{4k+2}
\left\{
\left(\frac{1}{1+x^{k+1}}\right)^3\cdot
\left(\frac{1}{1+x^{k+1}+m^2x^k}\right)^2
\right.\nonumber\\
&&\qquad\qquad\qquad\qquad\;\left.
+\left(\frac{1}{1+x^{k+1}}\right)^2\cdot
\left(\frac{1}{1+x^{k+1}+m^2x^k}\right)^3
\right\}\;,\\
I_3(k,m^2)&=&
\frac{k+1}{8\pi^2}
\int_0^\infty dxx^{3k+2}
\left\{
\left(\frac{1}{1+x^{k+1}}\right)^2\cdot
\left(\frac{1}{1+x^{k+1}+m^2x^k}\right)^2
\right.\nonumber\\
&&\qquad\qquad\qquad\qquad\;\left.
+2\left(\frac{1}{1+x^{k+1}}\right)^3\cdot
\left(\frac{1}{1+x^{k+1}+m^2x^k}\right)
\right\}\;,\\
I_5(k,m^2,\alpha)&=&
\frac{k^2-1}{8\pi^2}
\int_0^\infty dx\left[x^{2k+1}
\left\{
\left(\frac{1}{1+x^{k+1}}\right)^2\cdot
\left(\frac{1}{1+x^{k+1}/\alpha+m^2x^k}\right)
\right.\right.\nonumber\\
&&\qquad\qquad\qquad\qquad\quad\left.
+\left(\frac{1}{1+x^{k+1}}\right)\cdot
\left(\frac{1}{1+x^{k+1}/\alpha+m^2x^k}\right)^2
\right\}
\nonumber\\
&&\qquad\qquad\qquad
-x^{3k+2}
\left\{
\left(\frac{1}{1+x^{k+1}}\right)^2\cdot
\left(\frac{1}{1+x^{k+1}/\alpha+m^2x^k}\right)^2
\right.\nonumber\\
&&\qquad\qquad\qquad\quad\;\left.
+\left.2\left(\frac{1}{1+x^{k+1}}\right)^3\cdot
\left(\frac{1}{1+x^{k+1}/\alpha+m^2x^k}\right)
\right\}\right]\;,\\
I_4(k,m^2)&=&\frac{k+1}{k-1}I_5(k,m^2,\alpha=1).
\end{eqnarray}

Here we note some important features of  these
RG equations and integrals $I_x$.
The gauge parameter $\alpha$-dependence of the RG equations comes
solely from $I_5$, which vanishes with the special cutoff scheme
`$k=1$'. Therefore under this cutoff scheme,  
our RG equations are completely free from $\alpha$-dependence.
Of course this cancellation basically depends on the fact that 
our beta functions contain a gauge invariant set of diagrams, 
{\it e.g.}, the crossed box as well as the box diagrams, and 
inclusion of the proper anomalous dimension of fields.

The sharp cutoff (large $k$) limit makes the
RG equations singular. The integrals
$I_1$, $I_2$ and $I_3$ approach to finite functions of $m^2$, while 
$I_4$ and $I_5$ blow up to infinity, except for the case of 
vanishing photon mass ($m^2=0$). 
On the other hand, for $m^2=0$, integrals 
$I_4(k,m^2=0)$ and $I_5(k,m^2=0,\alpha)$ vanish independently of 
$k$ and $\alpha$. 
The photon mass $m^2$ is to be determined as a function of
the gauge coupling constant $e$ by the requirement of the
gauge invariance of the effective action 
$\Gamma_{\Lambda\rightarrow 0}$, 
and is the order of $e^2$. Therefore, the $\alpha$-dependence of 
our RG flow equations is the order of $e^4$.

We should note the essential difference of our NPRG flows
from the perturbative 
RG. Our RG equations describe the running  
of the 4-fermi and multi-fermi interactions as well, 
{\it e.g.}, $\partial G_S/\partial t\sim G_S^2$, which finally  gives
the effective potential of the composite operator $\bar\psi\psi$
so that we can
investigate the dynamical chiral symmetry breaking  within our RG
equations. Also the beta function of the gauge coupling constant $e$
contains the so-called quadratically divergent diagram contribution,  
$i.e.$, $\partial e/\partial t\sim eG_S$, which will need special care in relation to 
the gauge invariance.
%
%
\section{The critical behaviors of QED}

\hskip\parindent 
We investigate the $\GS-\GV$ sub-system described by 
the Eqs.(\ref{scalar}) and (\ref{vector}) with the vanishing photon mass
approximation ($m^2=0$). This system does not suffer from 
the gauge paremeter dependence, and also it has the infinite
$k$ limit which completely reproduces the
sharp cutoff LPA (Wegner-Houghton equation) results\cite{terao,teraow}. 
In more precise treatment with the 
running photon mass, a small gauge
dependence exists, except for the special cutoff profile of $k=1$.
The gauge coupling constant is fixed, no running. 
Complete analyses with the running gauge coupling constant will be
described in  the forthcoming paper \cite{FURTURE}. 
Our results here should be compared 
with the QED SD results with the fixed gauge coupling 
constant \cite{SDfourfermi}.
Then the RG equations determining the critical behaviors are 
the following coupled differential equations (note $I_3(k,0)=1/(8\pi^2)$):
\begin{eqnarray}
\frac{\partial}{\partial t}G_S&=&
-2G_S
+2I_1(k)G_S(3G_S+8G_V)
+\lambda G_S
-6\tilde{I_2}(k,0)\lambda^2,
\label{scalar2}\\
\frac{\partial}{\partial t}G_V&=&
-2G_V
-\lambda G_V
+I_1(k)G_S^2
-3\tilde{I_2}(k,0)\lambda^2,
\label{vector2}
\end{eqnarray}
where $\lambda$ and $\tilde{I_2}$ are defined by $\lambda\equiv3e^2/4\pi^2$ 
and $(4\pi^2/3)^2I_2$ respectively.

In the RG approach, the critical behaviors are governed by 
the structure of the fixed points. First we consider the case of 
vanishing gauge coupling constant, where the system is a pure
fermi system and we are solving the Nambu-Jona-Lasinio model
beyond the ladder (the large $N$ leading) approximation. As is
shown in Fig.\ref{njlflow}, we have
two fixed points, the Gaussian infrared fixed point at the origin
and the modified Nambu-Jona-Lasinio (NJL) ultraviolet fixed point
extended to the vector operator $\GV$
\footnote{To be precise, without gauge interactions, there is no
ultraviolet fixed point since the 8-fermi operators
do not have a fixed point solution.}. 
Also there appears another 3rd fixed point at the negative $\GS$ region,
which we will ignore here as a fake due to the approximation.

\vskip5mm
\refstepcounter{figure}
\label{njlflow}
\centerline{\epsfxsize=7cm\epsfbox{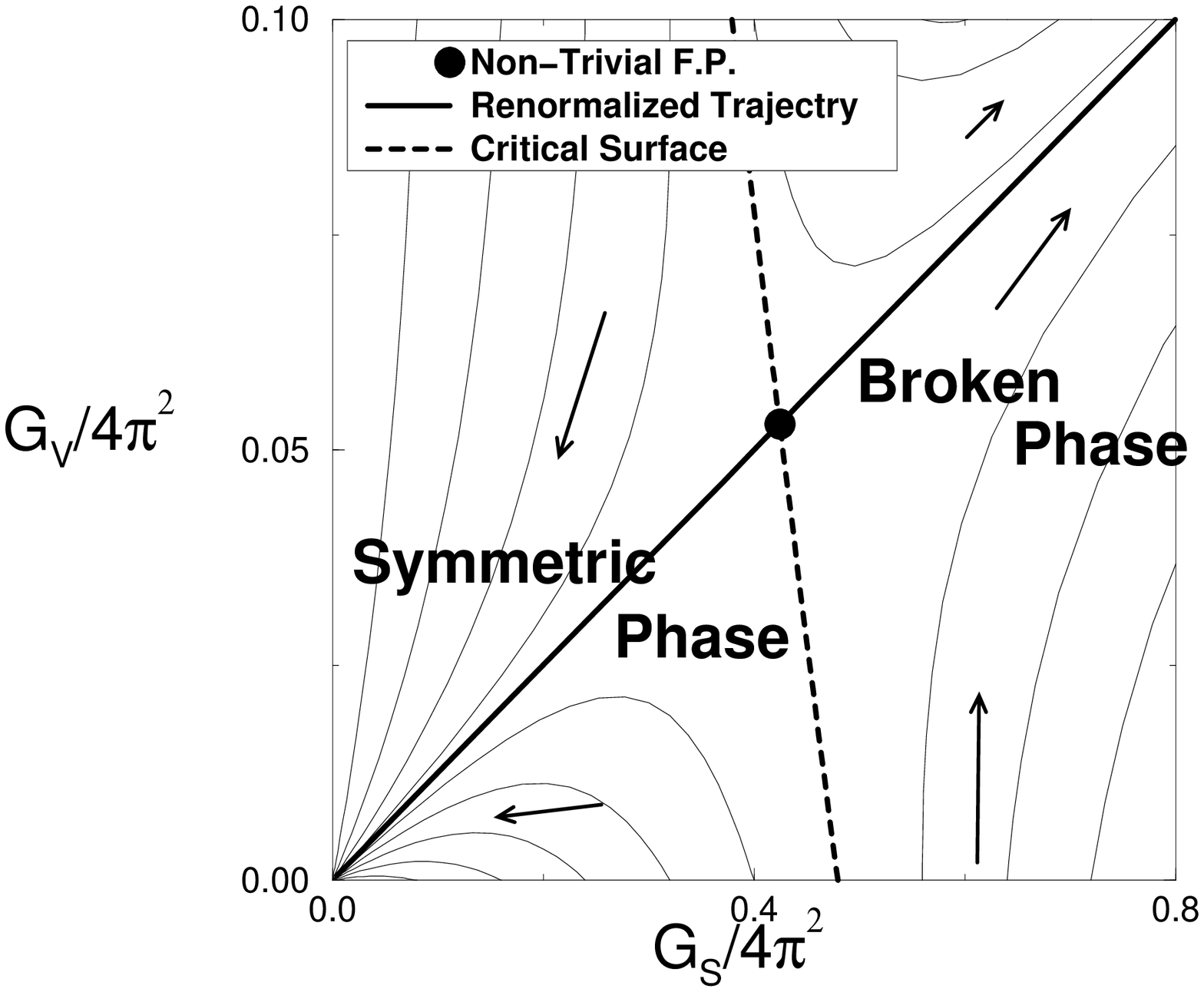}}
\vskip5mm
\centerline{\parbox{0.9\textwidth}{{\footnotesize
Figure \ref{njlflow}:
The flow diagram on the $G_S-G_V$ plane 
with the vanishing gauge
coupling constant and with the cutoff 
scheme `$k=1$'. 
}}}
\vskip8mm

There are two phases, the strong phase and the weak phase, and 
the renormalized trajectory is the straight line defined by $G_S=8G_V$. 
In the strong phase, the 4-fermi interactions grow up and 
diverge at the finite cutoff scale in this truncation. 
This divergence itself may be regarded as a signal of the limitation of our
truncation, 
and we expect that in an enlarged subspace this 
singularity will disappear. However, in the local 
potential approximation, we can still estimate
the dynamical fermion mass and the chiral condensate 
by taking account of the infinite number of 
higher dimensional fermi operators. In fact, investigating the
effective potential of the composite operator $\bar\psi\psi$ in either
phases, we can conclude that in the strong phase the chiral symmetry
is spontaneously broken while in the weak phase the chiral symmetry
is respected \cite{teraow}.

Switching on the gauge interactions (Fig.\ref{fpmove}), 
these two fixed points move
closer to each other and finally meet at some $\lambda$($=\LC$)
to pair-annihilate. Beyond this critical gauge coupling constant
$\LC$, no fixed point exists.
The Gaussian fixed point at $\lambda=0$ moves to some finite values of
$\GS, \GV$, which indicates the scale invariant 
infrared effective 4-fermi
interactions generated by the gauge exchange interactions.

\vskip5mm
\centerline{\epsfxsize=6cm\epsfbox{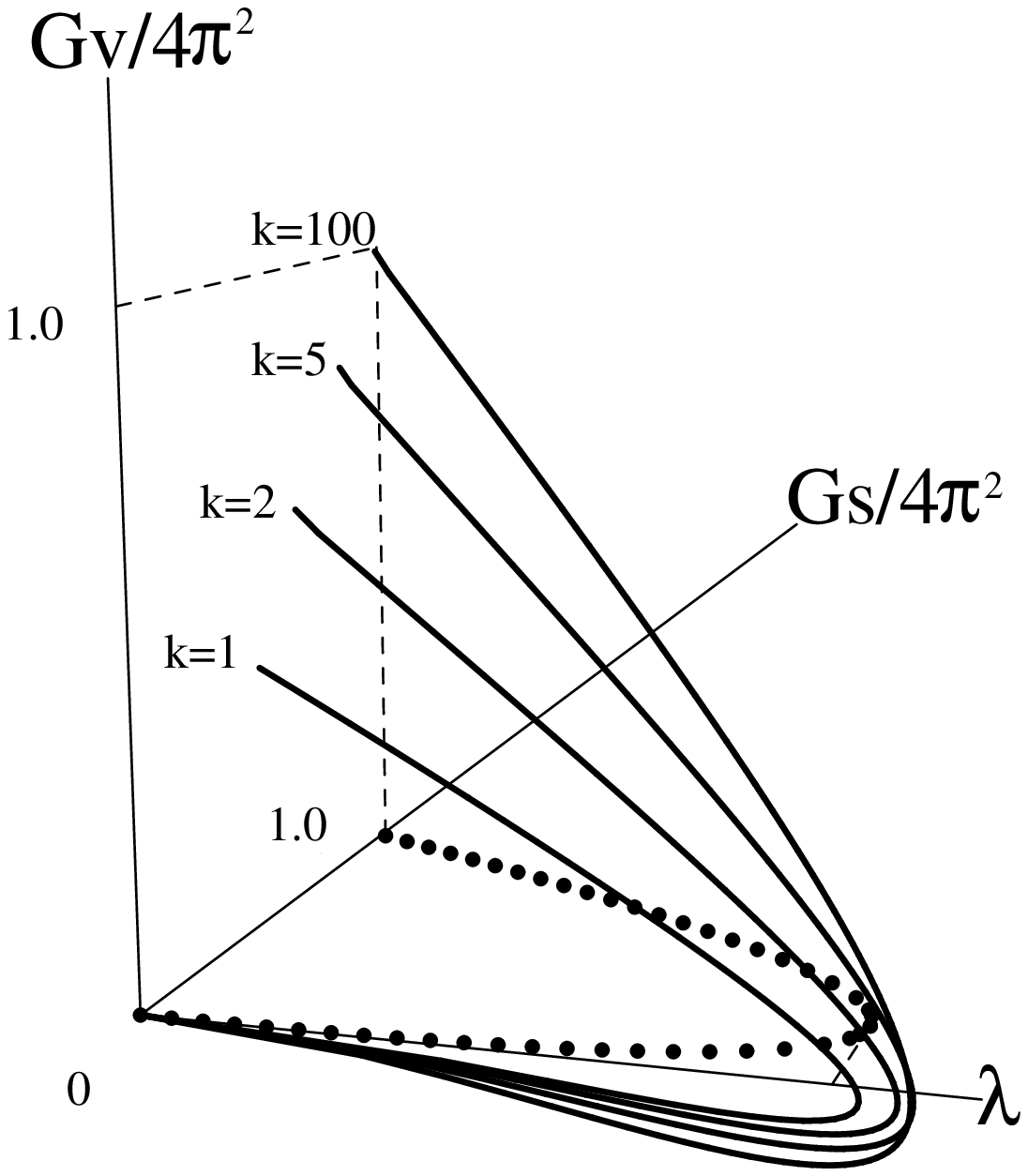}}
\vskip5mm
\refstepcounter{figure}
\label{fpmove}
\centerline{\parbox{0.9\textwidth}{{\footnotesize
Figure \ref{fpmove}:
The move of the fixed points with the various cutoff 
schemes. 
The ladder part LPA results\cite{terao,teraow}
giving the Landau gauge ladder SD equivalents 
are ploted by the dotted line.
}}}
\vskip8mm

We now understand the total phase structure. 
For the region $0\le\lambda < \LC$, there are 
two fixed points and two phases divided by the 
ultraviolet fixed point, the modified NJL
fixed point. For $\lambda > \LC$, no fixed point appears and the
whole theory space belongs to the strong phase of the dynamical 
chiral symmetry breaking. 
These structures generate the peculiar
phase diagram on the $\lambda - \GS $ plane. We
show the $k=1$ gauge independent results in Fig.\ref{phase}, where the
arrows show the RG flows. 

\vskip5mm
\centerline{\epsfxsize=7cm\epsfbox{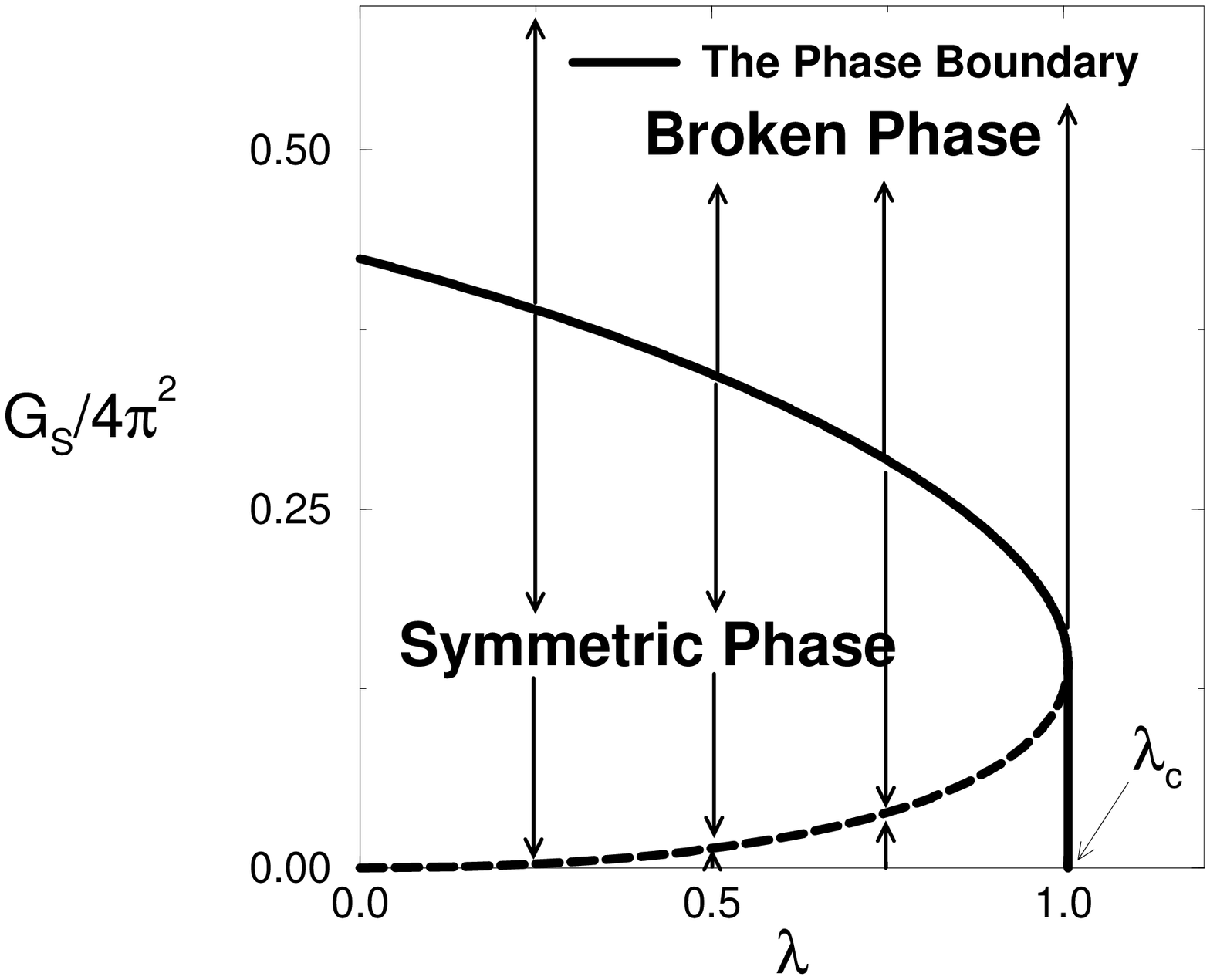}}
\vskip5mm
\refstepcounter{figure}
\label{phase}
\centerline{\parbox{0.9\textwidth}{{\footnotesize
Figure \ref{phase}:
The phase structure on the renormalized trajectory with the cutoff 
scheme `$k=1$'. We project it onto the $\lambda - \GS$ plane. 
Note that this is not equal to the section between 
the critical surface and the $\lambda - \GS$ plane which will be
shown in Fig.\ref{surface}
}}}
\vskip8mm

The critical surface obtained in various cutoff schemes
are drawn in Fig.\ref{surface}. In Figs.\ref{fpmove} and \ref{surface}, 
we see that the fixed point position and the critical surface depend
largely on the cutoff scheme.
This is seen in the RG equations where
the criticality is directly determined 
by the cutoff scheme dependent coefficients ($I_1$ and $I_2$). 
For example, the non-trivial fixed point on the $\lambda=0$ plane is given by 
$G_S^\ast=8G_V^\ast=1 /{(4I_1(k))}$. 
It should be mentioned that these quantities of the position of the
criticality are not the physical quantities at all, that is, 
they can not be measured.
Hence these large cutoff scheme dependences
do not cause any difficulty in our approach. 

\vskip5mm
\centerline{\epsfxsize=7cm\epsfbox{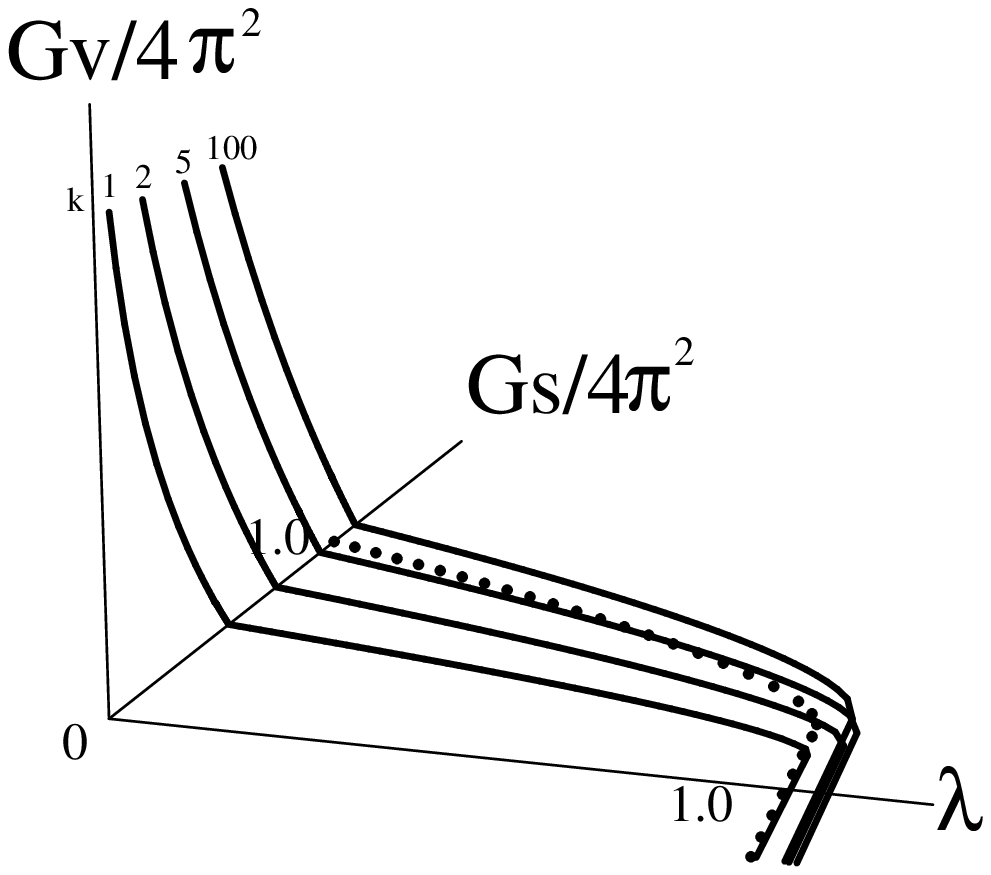}}
\vskip5mm
\refstepcounter{figure}
\label{surface}
\centerline{\parbox{0.9\textwidth}{{\footnotesize
Figure \ref{surface}:
The cutoff scheme dependence of the critical surface  
in the ($\GS,\GV,\lambda$) space. We show 
the cross sections of the critical surface with $\GS-\GV$ and
$\lambda - \GS$ planes. The dotted line is the result of the 
Landau gauge ladder SD equation. 
}}}
\vskip8mm

Our chiral phase criticality qualitatively coincides
with the Landau gauge ladder SD results
\cite{SDfourfermi}. The quantitative differences
come out of two sources. Our
results take account of the non-ladder contributions as well as the
ladder parts, which guarantees the gauge independence. 
Also our cutoff profile is smooth while the SD uses
the $\theta$-function sharp cutoff.
Taking a large $k$, say $k=100$, then our results are almost
equal to the LPA sharp cutoff results\cite{terao,teraow}
within the line width in every
figures.
%
%
\section{Critical exponent and anomalous dimension}

\hskip\parindent 
Here, we are going to evaluate the physical quantities, 
the critical exponent of the ultraviolet fixed point ($\nu_4$), and 
the anomalous dimension of the fermion mass operator ($\gamma_m$). 
The critical exponent $\nu_4$ is obtained by linearizing the
RG flow equations around the ultraviolet fixed point, and it is 
related to the anomalous dimension of the chiral invariant 4-fermi
operators $\gamma_4$ through $\nu_4 + 2 = \gamma_4$.
On the other hand, the fermion mass operator $\MF \bar\psi\psi$ 
explicitly breaks the chiral symmetry, and the vanishing mass 
point is its ultraviolet fixed point. Thus we may evaluate its
anomalous dimension $\gamma_m$ from the linearized RG 
equation for the operator $\MF \bar\psi\psi$, 
\begin{equation}
\frac{\partial}{\partial t}\MF=
\left (1+\gamma_m\right)\MF+O(\MF^3)\;.
\end{equation}
The anomalous dimension $\gamma_m$ is evaluated in terms of 
the gauge coupling constant $\lambda$ 
and the critical scalar 4-fermi interaction $\GS^\ast(\lambda)$ as 
\begin{equation}
\gamma_m(\lambda)=\frac{\lambda}{2}+8I_1(k)G_S^\ast(\lambda)\;,
\end{equation}
where the corresponding diagrams are shown in Fig.\ref{mass-fig}.

\vskip5mm
\centerline{\epsfxsize=14cm\epsfbox{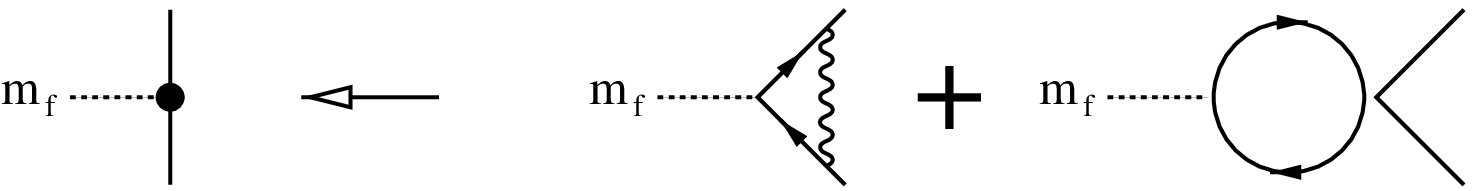}}
\vskip5mm
\refstepcounter{figure}
\label{mass-fig}
\centerline{\parbox{0.9\textwidth}{{\footnotesize
Figure \ref{mass-fig}:
Diagrams contributing to the anomalous dimension 
of the fermion mass operator $\bar\psi\psi$. 
}}}
\vskip8mm

\centerline{\epsfxsize=14cm\epsfbox{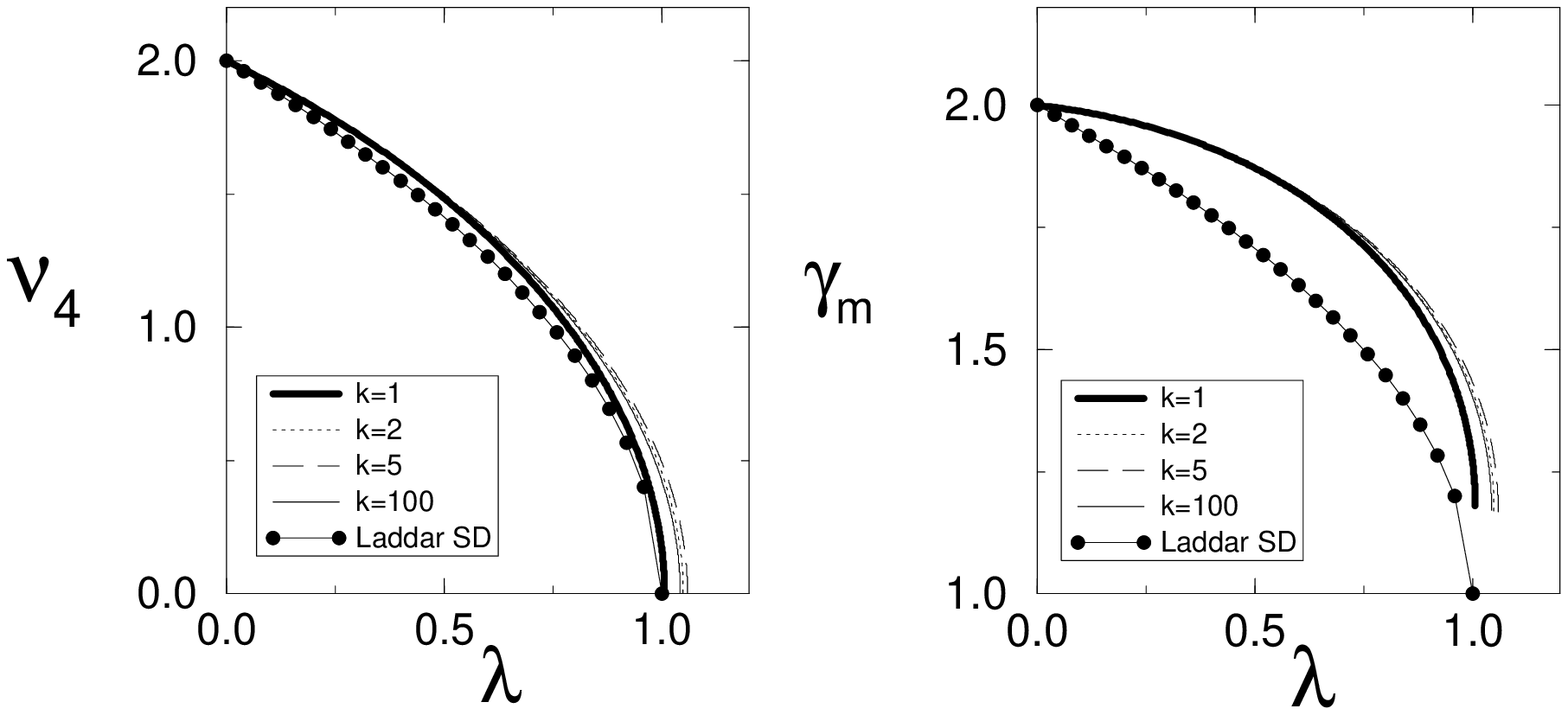}}
\vskip5mm
\refstepcounter{figure}
\label{fig6}
\centerline{\parbox{0.9\textwidth}{{\footnotesize
Figure \ref{fig6}:
The critical exponent $\nu_4$ 
and the anomalous dimension $\gamma_m$ in various cutoff schemes.
The dotted line is the result of the Landau gauge ladder SD equation. 
}}}
\vskip8mm

In Fig.\ref{fig6}, we plot the critical exponent $\nu_4$ and 
the anomalous dimension $\gamma_m$ as a function of 
the gauge coupling constant $\lambda$ obtained in various 
cutoff schemes, with those obtained by the ladder SD equation. 
Although there are large scheme dependence in the position of  criticalities, 
these exponent and anomalous dimension are 
almost independent of the cutoff scheme. This indicates that
our method and approximation are fairly stable against the cutoff
profile dependence.

Comparing with the ladder SD results, the anomalous dimension  $\gamma_m$ 
is enhanced much, while the critical exponent $\nu_4$ is almost the same.
This large corrections certainly come from 
the contribution of the non-ladder diagrams. 
Note that our results are gauge independent due to 
the inclusion of these non-ladder contributions.
In the ladder approximation, these two quantities are not independent and
satisfy a relation, $\nu_4+2 = 2\gamma_m$, which is 
given by the simple ladder relation of $\gamma_4 = 2\gamma_m$, where
only the scalar 4-fermi operators are taken into account.
In our non-ladder extension, this relation is broken and generates
the large deviation of $\gamma_m$.
The numerical values of our gauge independent results ($k=1$) 
are listed in
Table 1, comparing them with the Landau gauge ladder SD\cite{SDfourfermi} 
and the sharp cutoff LPA \cite{terao} results.
It should be
noted that the value of the critical gauge coupling constant 
$\LC$ depends on the 
approximation adopted.

\begin{table}
\begin{center}
\begin{tabular}{|c||c|c|c||c|c|c|} \hline
   & \multicolumn{3}{|c||}{$\nu_4$} & \multicolumn{3}{|c|}{$\gamma_m$} \\ 
\hline
$\lambda$ & ladder SD & LPA & our result & ladder SD & LPA & our result \\
 \hline \hline
0.0	&  2.000 &     2.000 &     2.000 &
  2.000 &     2.000 &     2.000 \\ \hline
0.1	&  1.897 &     1.918 &     1.918 &
  1.949 &     1.987 &     1.987 \\ \hline
0.2	&  1.789 &     1.828 &     1.827 &
  1.894 &     1.969 &     1.969 \\ \hline
0.3	&  1.673 &     1.729 &     1.726 &
  1.837 &     1.945 &     1.944 \\ \hline
0.4	&  1.549 &     1.619 &     1.614 &
  1.775 &     1.914 &     1.912 \\ \hline
0.5	&  1.414 &     1.497 &     1.487 &
  1.707 &     1.874 &     1.872 \\ \hline
0.6	&  1.265 &     1.359 &     1.342 &
  1.632 &     1.825 &     1.820 \\ \hline
0.7	&  1.095 &     1.200 &     1.173 &
  1.548 &     1.763 &     1.754 \\ \hline
0.8	&  0.894 &     1.013 &     0.968 &
  1.447 &     1.684 &     1.667 \\ \hline
0.9	&  0.632 &     0.776 &     0.699 &
  1.316 &     1.575 &     1.542 \\ \hline
1.0	&  0.000 &     0.418 &     0.177 &
  1.000 &     1.396 &     1.274 \\ \hline \hline
$\mbox{at } \LC$ &  0.000 &  0.000 &  0.000   & 
  1.000 &     1.170 &     1.177 \\ 
($\LC$)  &(1.0000)&(1.0409)&(1.0069) & (1.0000)&(1.0409)&(1.0069) \\ \hline
\end{tabular}
\vskip5mm
\parbox{0.9\textwidth}{{\footnotesize
Table 1. The critical exponent $\nu_4$ and the anomalous dimension $\gamma_m$. 
Our gauge independent results ($k=1$) are compared with the Landau
gauge ladder SD\cite{SDfourfermi} and the sharp cutoff
LPA\cite{terao}
results. 
}}
\end{center}
\label{table1}
\end{table}

\section{Discussions and Comments}

\hskip\parindent 
We solved the next to LPA non-perturbative renormalization 
group and obtained the critical behaviors of the dynamical 
chiral symmetry breaking in QED.
Our RG system with a special cutoff scheme ($k=1$) exhibits 
the gauge parameter independence, which is assured by the
inclusion of the non-ladder contributions like the crossed box diagrams.
The NPRG approach to the dynamical chiral symmetry breaking in gauge theories
has great potentiality. It turns out that the NPRG with our 
subspace approximation gives a
beyond the ladder approximation restoring the gauge independence.
It should be noted that the $k=1$ cutoff scheme does not work if
we go to the higher orders of the derivative expansion, since the
beta function integrals do not converge. However, we expect that
the gauge dependence should not be large since the NPRG approach may easily
incorporate the `gauge invariant' set of diagrams.

We have also confirmed the stability of our method. The cutoff scheme
dependences in physical quantities are found to be negligible.
Also we get rather small deviations of physical results
compared to the LPA. For example, the anomalous dimension $\gamma_m$
at $\LC$ is 1.170 in the LPA and 1.177 in our next to LPA respectively.
In our level of approximation in this article ($m^2=0$), this
stability is actually equivalent to the cutoff scheme independence since
the infinite $k$ limit corresponds to the sharp cutoff LPA.

\vskip5mm
\noindent
{\bf Acknowledgment}
\vskip2mm

One of the authors(H.~T.) thanks the Department of Physics, University of Genova 
for hospitality while this paper was completed. 
K-I.~A.~ and H.~T. are supported in
part by Grant-in Aid for Scentific Research
(\#08240216 and \#08640361) from the Ministry of Education,
Science and Culture.

\end{document}